\documentclass[aps,prl,twocolumn,superscriptaddress,floatfix]{revtex4-1}
\usepackage{blindtext}
\usepackage{amsmath}

\usepackage{hyperref}
\hypersetup{
  colorlinks   = true, 
  urlcolor     = blue, 
  linkcolor    = blue, 
  citecolor   = blue 
}

\usepackage{bm}
\usepackage{gensymb}
\usepackage[normalem]{ulem}
\usepackage{euscript}
\usepackage[pdftex]{graphicx}
\usepackage{xspace}
\usepackage{xfrac}
\usepackage{enumerate}
\usepackage{braket}
\usepackage{float,fancyvrb}
\usepackage[T1]{fontenc}
\usepackage{lmodern}
\graphicspath{{./Figures/}}
\usepackage{wrapfig}
\usepackage{scrextend}
\usepackage{comment}
\usepackage{url}
\usepackage{amssymb}
\usepackage{xcolor}

\def\be{\begin{equation}}
\def\ee{\end{equation}}
\def\beq{\begin{eqnarray}}
\def\eeq{\end{eqnarray}}
\def\ba#1{\begin{array}{#1}}
\def\ea{\end{array}}
\def\bn{\begin{enumerate}}
\def\en{\end{enumerate}}

\definecolor{ao}{rgb}{0.0, 0.5, 0.0}

\begin{document}

\title{Dynamical obstruction to localization in a disordered spin chain}

\author{Dries Sels}
\affiliation{Department of Physics, New York University, New York, NY, USA}
\affiliation{Center for Computational Quantum Physics, Flatiron Institute, New York, NY, USA}
\affiliation{Department of Physics, Harvard University, Cambridge MA, USA}
\author{Anatoli Polkovnikov}
\affiliation{Department of Physics, Boston University, Boston, Massachusetts, USA}
\date{\today}

\begin{abstract}
We analyze a one-dimensional XXZ spin chain in a disordered magnetic field. As the main probes of the system's behavior we use the sensitivity of eigenstates to adiabatic transformations, as expressed through the fidelity susceptibility, in conjunction with the low frequency asymptotes of the spectral function. We identify a region of maximal chaos -- with exponentially enhanced susceptibility -- which separates the many-body localized phase from the diffusive ergodic phase. This regime is characterised by slow transport and we argue that the presence of such slow dynamics is incompatible with the localization transition in the thermodynamic limit. Instead of localizing, the system appears to enter a universal subdiffusive relaxation regime at moderate values of disorder, where the spectral function of the local longitudinal magnetization is inversely proportional to the frequency, corresponding to logarithmic in time relaxation of its auto-correlation function. 
\end{abstract}

\maketitle

\section{Introduction}

Searching for non-ergodic/non-thermalizing systems, stable to perturbations, has been a long quest since the seminal work by Fermi, Pasta,  Ulam, and Tsingou~\cite{berman2005FPU, porter2009FPU}. In low dimensional setups it is known that such systems exists due to KAM theorem~\cite{arnold1989mechanics}. However, in extended systems whether classical or quantum, existence of such stable non-ergodic systems largely remains an open problem. There are several known situations when integrable systems remain perturbatively stable, i.e. they do not thermalize within any order of perturbation theory in a small parameter. Just to name a few examples we mention periodically driven (Floquet) systems, which thermalization time is exponential (non-analytic function) either in the inverse frequency~\cite{abanin2015heating,mori2016heating, weinberg2017APT, rajak2018classical_floquet, howell2019prethermalization} or to the amplitude of perturbation~\cite{vajna2018replica, roeck2019slow}. Numerically it was found that certain Flqouet quantum systems are exceptionally robust against heating even in the absence of obvious small parameters~\cite{prosen1998kicked_spins, dalessio2013floquet_localization, haldar2018floquet, heyl2019trotter}, however it is still not known what is their fate in the thermodynamic limit. All the models analyzed in these papers are perturbatively stable. In two seminal papers~\cite{gornyi2005MBL, basko2006MBL} it was argued that disordered systems in the insulating regime are perturbatively stable against adding small short range interactions. As such they should have strictly zero conductivity at finite temperatures in the thermodynamic limit. This phenomenon was termed many-body localization (MBL). In Ref.~\cite{oganesyan2007localization} it was proposed that MBL is stable beyond the perturbative regime in one-dimensional systems and therefore the MBL phenomenon might persist in thermodynamic limit. In Ref.~\cite{imbrie2016MBL} J. Imbrie proposed a proof for existence of MBL regime in a disordered Ising model using some extra assumptions. After the initial discovery, MBL attracted a lot of attention both theoretical and experimental and we refer to recent reviews~\cite{rahul2015,abanin2019review} for further information and references.

Recently, several numerical studies have raised doubts about the stability of a localized phase in the thermodynamic limit. Systems with quasi-periodic potentials have been argued to be unstable against interactions by {\v Z}nidari{\v c} and Ljubotina~\cite{Znidaric}. Two recent papers by J. \~Suntajs et al.~\cite{suntajs2019quantum, suntajs2020transition} provide numerical evidence for the drift of the localization transition point with the system size towards infinitely large disorder strength in the thermodynamic limit, by studying the spectral form factor. These findings were challenged in a number of works, including Refs.~\cite{abanin2019rebuke,PhysRevLett.124.186601,Panda_2020}. The main critique being that one can not reliably extrapolate the information obtained on small systems to the thermodynamic limit. Finally, Kiefer-Emmanouilidis et al.~\cite{PhysRevLett.124.243601,kieferemmanouilidis2020absence} provide evidence for slow growth of the number entropy at strong disorder; critique on~\cite{PhysRevLett.124.243601} was formulated in Ref.~\cite{PhysRevB.102.100202}, arguing that no transport is associated with the increasing entropy. One of the goals of the present paper is to analyze the stability of the localized phase to increasing the system size using the sensitivity of eigenstates to infinitesimal local perturbations and the low frequency behavior of the spectral function. Combining both measures, we provide further arguments for the idea that there is no truly localized phase in the thermodynamic limit in one dimensional disordered systems, at least in the particular disordered XXZ model we are studying here. These conclusions are supported by a general argument that proposed earlier Griffith type scenarios of the MBL transition are inconsistent with the spectral sum rule in generic local interacting models. The ultimate physical reason behind this conclusion is the existence of an intermediate maximally chaotic region, found earlier in clean systems~\cite{pandey2020chaos} and a disordered central spin model~\cite{villazon2020CNE}, which separates the localized and ergodic regimes in finite systems and prevents the system from localization in the thermodynamic limit.

In the last decade it was realized that fidelity susceptibility $\chi$, or more generally the quantum geometric tensor, is a very efficient measure for detecting zero-temperature quantum phase transitions (see for example Refs.~\cite{venuti2007geometrictensor, albuquerque2010fidelity, kolodrubetz2013geometry, kolodrubetz2017geometry}). Fidelity susceptibility defines the sensitivity of the (ground) state to small perturbations. Near the phase transition this sensitivity is usually amplified, leading to singular behavior of $\chi$; often a divergence. Physically $\chi$ is determined by the low-frequency tail of the spectral function, which is enhanced near phase transition boundaries due to the critical slowing down.  At finite energy densities the utility of the fidelity susceptibility was less obvious because of its exponential divergence with the system size in ergodic systems~\cite{kolodrubetz2017geometry}. We note that the there are finite-temperature generalizations of the fidelity susceptibility measuring distance between density matrices like the Bures metric, which in turn is related to the Fisher information (see e.g. Refs.~\cite{sidhu2020bures, liska2020hidden}). However, the latter are not sensitive to chaos or integrability rather probing equilibrium thermodynamic properties of the system. 

In recent work we have shown that the he norm of the adiabatic gauge potential (AGP), which is equivalent to the fidelity susceptibility $\chi_n$ averaged over different eigenstates $n$, can serve as a very sensitive probe of quantum chaos~\cite{pandey2020chaos}. The AGP norm is able to pick up tiny (exponentially small in the system size) integrability breaking perturbations, which are not necessarily visible to traditional measures of quantum chaos like the level repulsion~\cite{d2016quantum}, the spectral form factor~\cite{suntajs2019quantum}, or the survival probability~\cite{borgonovi2016quantum}. This sensitivity comes from the fact that the low frequency tail of the spectral function defining $\chi_n$ can detect changes at very (exponentially) long time scales, where small perturbations should leave the most pronounced effect. In particular, dependence of the AGP norm on the system size changes from polynomial for integrable systems to exponential for chaotic systems and this change is very easy to detect numerically. One of the key findings of that work was that the transition from the integrable to ergodic phase across various models happens through an intermediate phase, where the fidelity susceptibility diverges even faster than in the ergodic regime and this divergence is accompanied by exponentially long in the system size relaxation times. In another recent work~\cite{villazon2020CNE}, based on a similar analysis, it was further argued that in a particular disordered central spin model this chaotic behavior coexists with nonthermalizing nature of the individual eigenstates, and hence to non-ergodic behavior of the system even in the absence of small parameters. 

In this work we adopt the approach of Ref.~\cite{pandey2020chaos} to analyze properties of a disordered one-dimensional XXZ spin chain and specifically analyze the fate of the many-body localization transition and the nature of the non-ergodic regime there. We use the standard Hamiltonian
\be
H= \frac{1}{W} \sum_j ( S^x_j S^x_{j+1}+S^y_j S^y_{j+1}+\Delta S^z_j S^z_{j+1})  + \sum_j h_j S_j^z,
\ee
where $S_j^{x,y,z}$ are the spin 1/2 operators and $h_j$ are uncorrelated random numbers uniformly distributed in the interval $[-1,1]$. We assume periodic boundary conditions and we fix the anisotropy parameter at $\Delta=1.1$ such that the model is close to the extensively studied Heisenberg spin chain~\cite{serbyn2015criterion, suntajs2019quantum} and at the same time has broken $SU(2)$ symmetry even in the absence of disorder. It is expected that these minor modifications of the model do not affect any results related to the MBL transition apart from a small shift of the critical disorder strength.

\section{Fidelity Susceptibility}

The fidelity susceptibility, or equivalently the diagonal component of the quantum geometric tensor, with respect to some coupling $\lambda$ of a given eigenstate $n$ is defined as~\cite{venuti2007geometrictensor, kolodrubetz2017geometry}
\be
\label{eq:chi_n}
\chi_n=\langle n|\overleftarrow{\partial_\lambda} \partial_\lambda|n\rangle_c\equiv  \sum_{m\neq n}{|\langle n |\partial_\lambda H |m\rangle|^2\over (E_n-E_m)^2}
\ee
The Hilbert-Schmidt norm of the AGP is defined as an average over the eigenstates of $\chi_n$. In order to avoid large eigenstate to eigenstate fluctuations, but keep exponential sensitivity of this probe, it was proposed in Ref.~\cite{pandey2020chaos} to additionally regularize $\chi_n$ by introducing an energy cutoff $\mu$ that is exponentially small in the system size. 

As a probe $\lambda$ we will use the local longitudinal magnetic field acting on a single spin, i.e. 
\be
H\to H+\lambda S_l^z,\quad \partial_\lambda H=S_l^z
\ee
All calculations will be done at $\lambda=0$, i.e. we will analyze sensitivity of eigenstates to an infinitesimal increase of a Z-magnetic field on a single site. By direct inspection it becomes clear that the disorder average fidelity susceptibility exponentially diverges with the system size as long as $W < \infty$. At large $W$, this divergence comes from rare resonances where $|E_m - E_n|\ll 2^{-N}$, which dominate the average $\chi_n$ in the absence of level repulsion. In contrast, since the resonances are rare, the typical fidelity susceptibility does not diverge with the system size at sufficiently large $W$. As such, it should become small since eigenstates are almost polarized along the Z-axis and adding an extra magnetic field does not affect them. For this reason, it is much more convenient to analyze the scaling behavior of the typical $\chi_n$ obtained by averaging its logarithm. We will define the typical log-fidelity susceptibility as
\be
\zeta=\langle\langle\,\log(\chi_n)\,\rangle\rangle,
\ee
where $\langle\langle\dots\rangle\rangle$ stands for averaging over both different disorder realizations and different eigenstates. We note that, in the ETH regime, $\zeta$ is equivalent to the logarithm of the AGP norm analyzed in Ref.~\cite{pandey2020chaos} since the susceptibility is concentrated around the mean (see e.g. Fig.~\ref{fig:distributionAGP} A). 

\begin{figure}[ht]
	\centering
	\includegraphics[width= 0.48\textwidth]{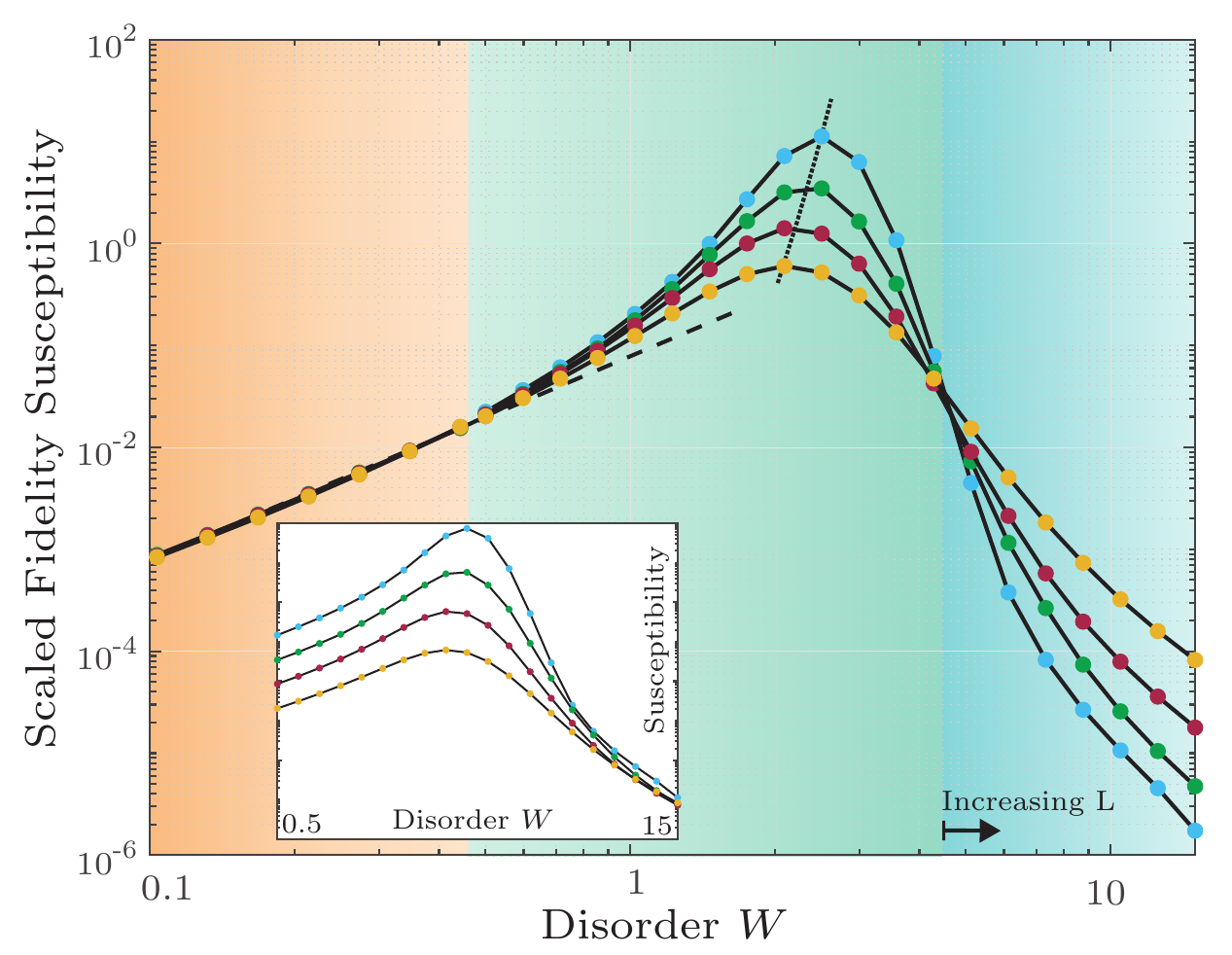}
	\caption{\textbf{Typical scaled fidelity susceptibility:} $\exp[\zeta]/2^L$ (main plot) and the unscaled one: $\exp[\zeta]$ (inset) as a function of disorder for different system sizes $L=12, 14, 16, 18$ (yellow to blue). The dashed line shows the ETH prediction. The dotted line shows the predicted peak drift with the system size: $\chi_{\rm max}=c\exp[\alpha W^\ast]$, where $W^\ast=L \log(2)/\alpha$ (see Eq.~\eqref{eq:W_ast}). Yellow, green and blue shaded regions correspond to ergodic, glassy and localized phases. As the system size increases the green blue crossover boundary drifts to stronger disorder values. }
\label{fig:typicalAGP}
\end{figure}

It is a simple exercise to extract asymptotes of $\zeta$ at weak and strong disorder. At small $W$ the system is expected to be ergodic and fully obey ETH~\footnote{We exclude very small values of disorder $W\lesssim 0.1$ where the system becomes very close to the integrable XXZ model}. Then, see e.g. Ref.~\cite{pandey2020chaos}, one expects
\be
\label{eq:zeta_eth}
\zeta = L\log 2+A,
\ee
The constant $A$ generally depends on the couplings in the Hamiltonian but is insensitive to the system size (up to possible logarithmic corrections). In the opposite regime of strong disorder it is easy to show using perturbation theory in $1/W$ that (see Appendix~\ref{app:Scaling})
\be
\zeta= B-{8\over 3} \log W,
\ee
where $B$ is another purely numerical constant, which neither depends on the system size nor on the couplings. In Ref.~\cite{pandey2020chaos} we found that in non-disordered models there is a new robust intermediate regime separating integrable and ergodic limits, where
\be
\label{eq:zeta_max}
\zeta = 2 L\log 2+A',
\ee
and the constant $A'$ is also independent of the system size. Such scaling of $\zeta$ saturates its upper bound. Physically, it corresponds to presence of exponentially slow in the system size relaxation times in the system or equivalently exponential concentration of the spectral weight at frequencies of the order of the level spacing (see next section for more details).

In Fig.~\ref{fig:typicalAGP} we show $\zeta$ as a function of the disorder strength $W$ for different system sizes $L=12,14,16,18$.  The main plot shows the fidelity scaled by the expected ETH asymptote: $\exp[\zeta]/2^L$, while the inset shows the unscaled susceptibility $\exp[\zeta]$. The small and large disorder behavior of $\zeta$ agree with general expectations. Thus for $W<1/2$ all curves collapse into the expected ETH prediction shown in dashed line. It is easy to understand the dependence of the constant $A$ in Eq.~\eqref{eq:zeta_eth} on disorder from the dimensional analysis: $A\propto W^2$. Likewise from the inset it is evident that at large disorder all curves become system size independent perfectly with the slope close to $8/3$, in the perfect agreement with the perturbation theory (see Appendix\ref{app:Scaling}). In the intermediate disorder regime $0.5\lesssim W \lesssim 10$ we see clear deviations from both asymptotes. In particular, there is a clear maximum in $\exp[\zeta]$ developing at intermediate values of disorder $W\in [2,4]$. This maximum drifts to higher values of disorder as we increase the system size, gradually approaching the upper bound~\eqref{eq:zeta_max}. As we discussed above, such scaling implies exponentially long in the system size relaxation times and correspondingly exponentially large values of the spectral weight at frequencies of the order of the level spacing. Simultaneously with the drift of the maximum to higher $W$ with increasing $L$, the curves tend to sharpen and the main plot even suggests that there could be a transition to a new phase (MBL) at disorder $W\sim 4.5$. However, as we discuss later, this conclusion would be premature and a large part of this work will be devoted to understanding that drift.
In the next section we will analyze what goes on by directly looking into the spectral function.

\section{Spectral Function}

To get more insight into the behavior of the system we analyze the spectral function, whose low frequency nature sets the scaling of the fidelity susceptibility~\cite{pandey2020chaos}. Our work is by no means the first work to look into the dynamical behavior of disordered systems through linear response probes. Ample works have numerically investigated the two-point spin correlator as well as the optical spin conductivity~\cite{Kratiek2015,Assa2016,luitz2017subdiffusion,PhysRevLett.117.040601,
PhysRevB.90.064203,PhysRevB.94.045126,PhysRevB.96.104201}, either through real time dynamics or in Fourier space.
This spectral function is defined as a Fourier transform of the symmetric correlation function of the local magnetization $S_l^z\equiv \partial_\lambda H$ (to simplify notations we suppress the site index $l$ in the notation for the spectral function). In this paper we will only consider the spectral function averaged over eigenstates:
\be
\label{eq:spectral_function_def}
|f(\omega)|^2={1\over \mathcal D}\sum_n\int_{-\infty}^\infty \frac{dt}{2\pi}\,\mathrm e^{i\omega t} G^n_z(t),
\ee
where $\mathcal D=2^L$ is the dimension of the Hilbert space and $G^n_z$ is the connected correlation function:
\[
G^n_z(t)\equiv 
\frac{1}{2} \langle n| S_l^z(t) S_l^z(0)+ S_l^z(0) S_l^z(t) |n\rangle_c
\]
with
\[
 \langle n|S_l^z(t) S_l^z(0)|n\rangle_c\equiv  \langle n|S_l^z(t) S_l^z(0)|n\rangle-\langle n| S_l^z|n\rangle^2.
\]
The spectral function satisfies a simple f-sum rule, which immediately follows from integrating Eq.~\eqref{eq:spectral_function_def} over frequency $\omega$:
\begin{eqnarray}
\label{eq:sum_rule}
\int\limits_{-\infty}^\infty d\omega\,|f(\omega)|^2&=&{1\over \mathcal D}\sum_n\langle n |(S_l^z)^2 |n\rangle_c \nonumber \\
&=&{1\over 4\mathcal D}\sum_n \left (1-4 \langle n| S_l^z|n\rangle^2 \right),
\end{eqnarray}
which is nothing but the average (over eigenstates) fluctuations of the local magnetization, which are in turn equivalent to the averaged non-conserved (decaying) part of the local magnetization. In the MBL phase it is expected that part of the local magnetization is conserved, implying that $0<\langle n |S_l^z|n\rangle_c<1/4$. In contrast, in the ergodic phase  $\langle n |(S_l^z)^2|n\rangle_c\to 1/4$ as $L\to \infty$. 

\begin{figure}[ht]
	\centering
	\includegraphics[width= 0.48\textwidth]{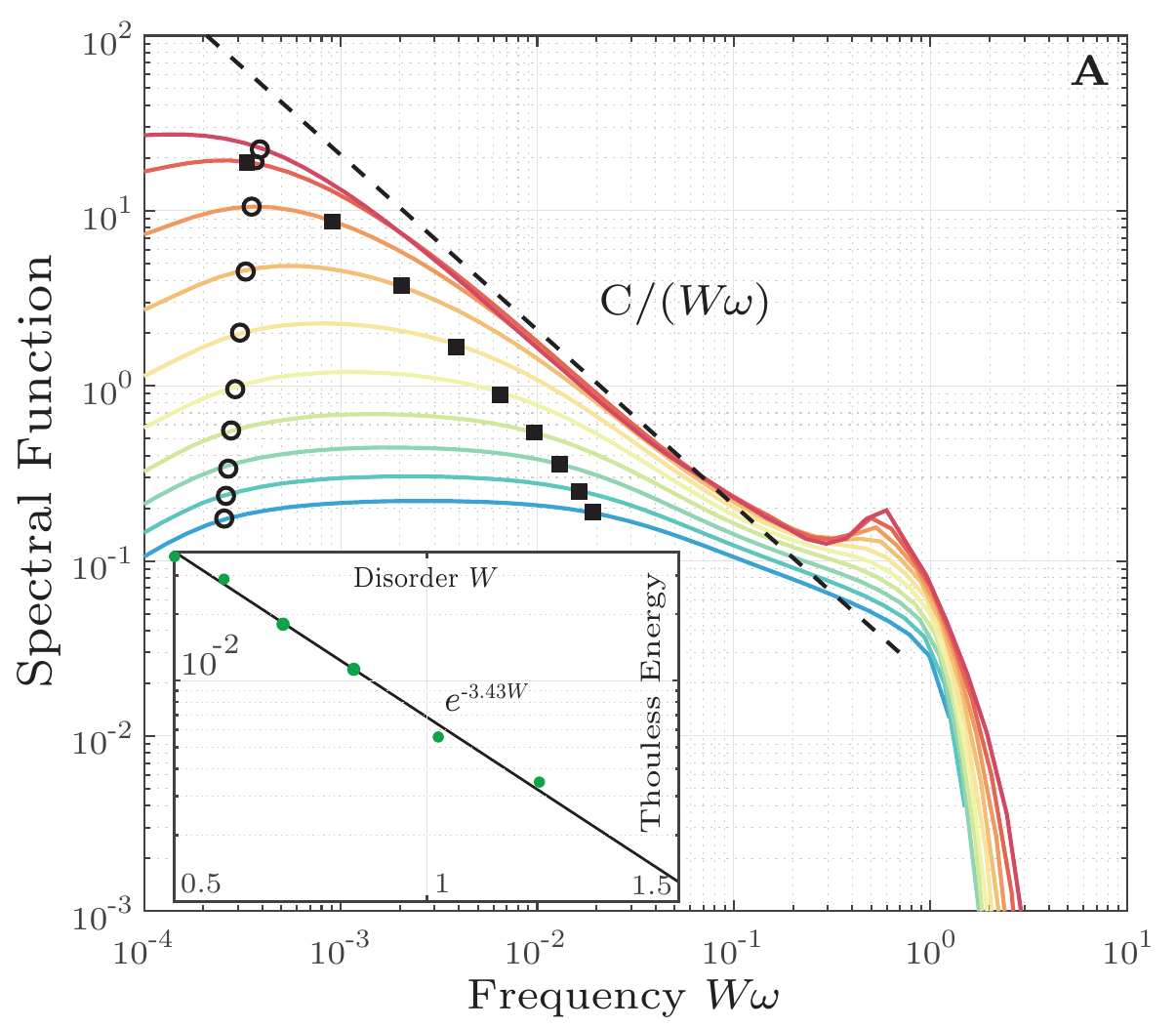}
	\includegraphics[width= 0.48\textwidth]{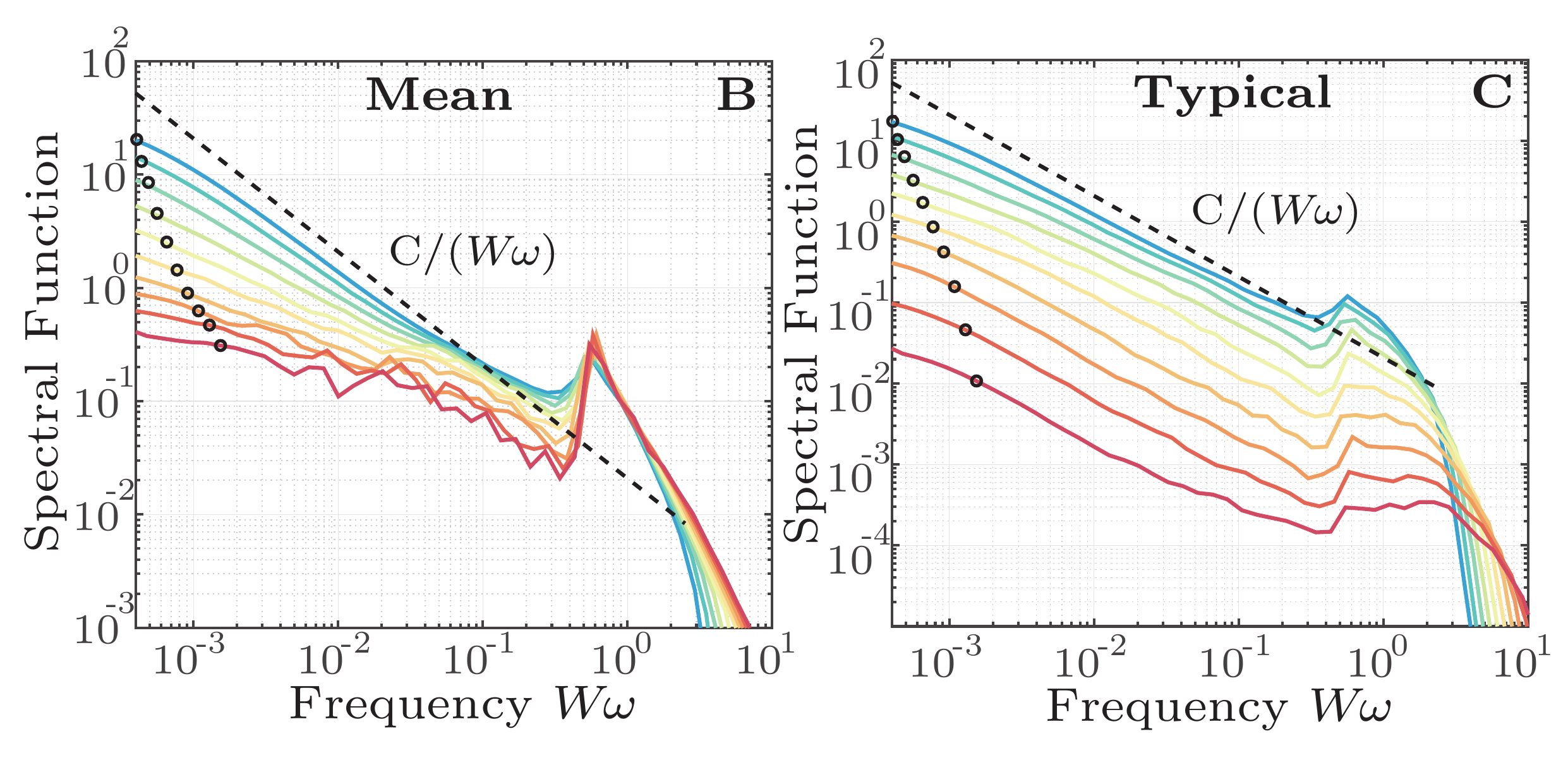}
	\caption{\textbf{Spectral function I.} The spectral function $|f(\omega)|^2$ defined according to Eq.~\eqref{eq:spectral_function_def} is shown for various values of the disorder $W$ in a $L=16$ chain. {\bf Panel A} Shows disorder increasing logarithmically, i.e. equally spaced, from $W=0.5$ (blue) to $W=2.5$ (red). The level spacing, or inverse Heisenberg time, is indicated by the black circles and an estimate of the Thouless energy is shown by the black squares~\cite{NoteFits}. The inset shows an exponential fit of the Thouless energy as a function of disorder. {\bf Panel B} Shows the average spectral function with disorder increasing logarithmically equally spaced from $W=2.5$ (blue) to $W=15$ (red). The typical level spacing is again shown by the black circles. {\bf Panel C} Instead  shows the typical spectral function, defined by averaging $\log|f(\omega)|^2$ over different disorder realizations, for the same disorder values as in panel B. The dashed line is a guide for the eye and shows asymptotic inverse frequency scaling of the spectral function: $|f(\omega)|^2=C/(W \omega)$. }
\label{fig:spectralL16}
\end{figure}

\begin{figure}[ht]
	\centering
	\includegraphics[width= 0.48\textwidth]{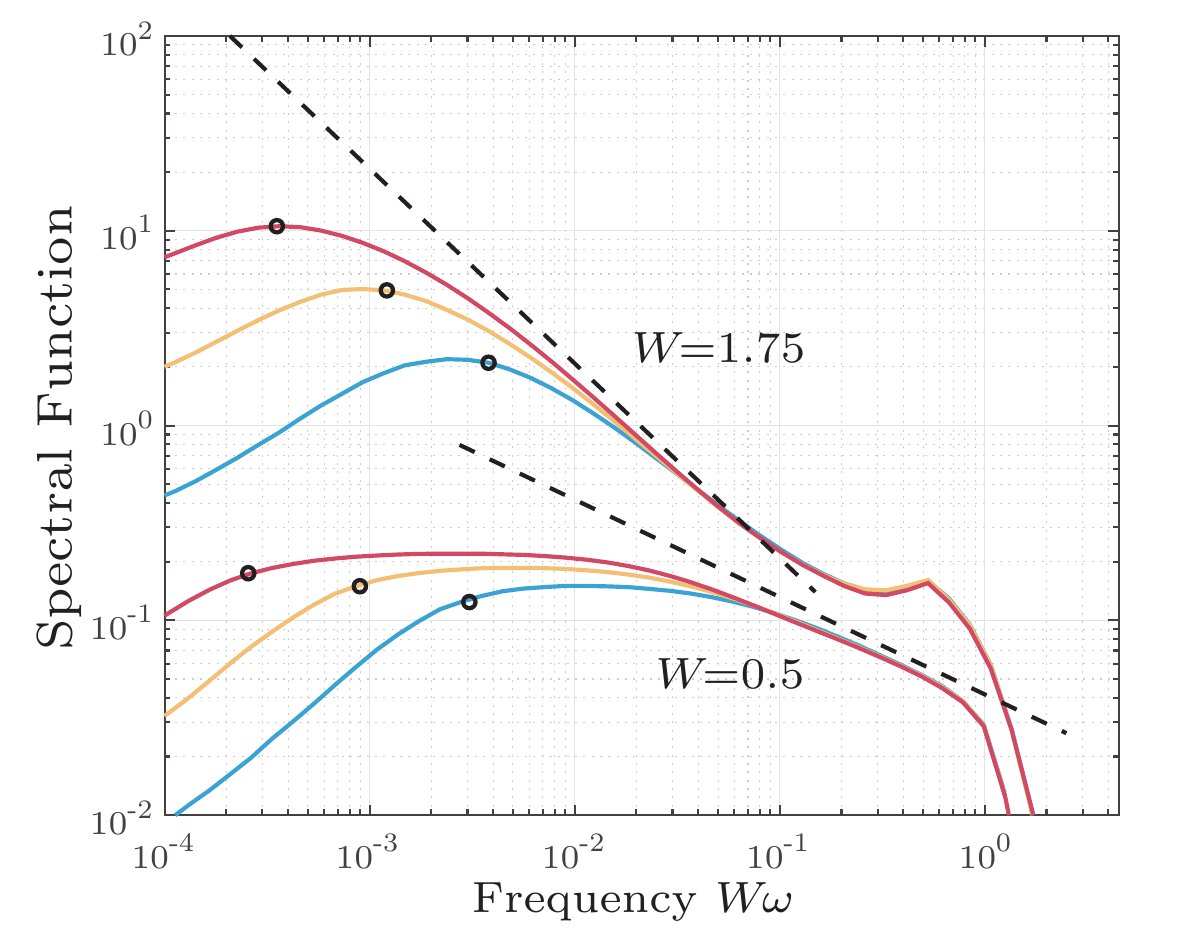}
	\caption{\textbf{Spectral function II.} The Fourier transform of the infinite temperature connected two-point correlation of the local $z-$magnetization is shown for two values of the disorder $W=0.5$ and $W=1.75$ for chains of size $L=12,14,16$. Heisenberg time is indicated by the black circles and the black squared present an estimate of the Thouless time~\cite{NoteFits}. The black dashed lines show $1/\sqrt{\omega}$ and $1/\omega$ asymptotes.}
\label{fig:spectralLscaling}
\end{figure}

The top panel of Fig.~\ref{fig:spectralL16} shows the averaged over disorder realizations spectral function $|f(\omega) |^2$ for low to intermediate values of disorder which incrementally changes from $W=0.5$ (bottom blue line) to $W=2.5$ (top red line). Note that all these values of disorder are in the delocalized regime corresponding to the shaded green color in Fig.~\ref{fig:typicalAGP}. One can observe a clear crossover from diffusive to subdiffusive behavior of the magnetization. In the diffusive regime the correlation function $G^n_z(t)$ decays as~$\sim 1/\sqrt{t}$ at long times, which is reflected in the  $|f(\omega)|^2\sim1/\sqrt{\omega}$ scaling of the spectral function, which eventually saturates at low frequencies below the inverse Thouless time. Such diffusive scaling is clearly seen at $W=0.5$. As disorder increases the spectral function crossovers to $1/\omega$ asymptote, which corresponds to much slower, logarithmic in time decay of the local magnetization: $G_z(t)~\sim A- B\log(t)$ with A and B constants. 

A sharp transition from diffusion to subdiffusion was previously reported in Ref.~\cite{PhysRevLett.117.040601} based on a study of the stationary state of a large (400 spin) boundary driven XXZ chain. Subdiffusion at these values of disorder was also reported before in Refs.~\cite{Kratiek2015,Assa2016,luitz2017subdiffusion,PhysRevB.96.104201} with a varying dynamical exponent as a function of $W$, which diverges at some critical disorder attributed to the MBL transition. The latter support a Griffiths phase interpretation~\cite{kratiekreview} of the ergodic side of the transition. Recent works however suggest that the subdiffusion might be transient~\cite{PhysRevLett.118.196801,PhysRevB.100.214313} and ordinary diffusion might govern the late time behavior of the system in the thermodynamic limit.

Although no firm statements can be made based on the present numerics, the data might support the latter. At weak disorder $W<0.5$ we observe a stable $1/\sqrt{\omega}$ regime which becomes flat at the Thouless energy and whose exponent doesn't drift with system size (see Fig. Fig.~\ref{fig:spectralLscaling}). At larger disorder it is much harder to identify a stable power law regime and the spectral function clearly has a smaller exponent at smaller frequencies. Moreover, at intermediate frequencies the exponent increases with the system size $L$ approaching one, as can seen from Fig.~\ref{fig:spectralLscaling}, where we show the spectral function for two different realization of disorder: $W=0.5$ and $W=1.75$ and three different system sizes: $L=12, 14, 16$.

We thus propose the following interpretation of the data, beyond a critical value of disorder $W\sim 0.5$, corresponding to the border between yellow and green regions in Fig.~\ref{fig:typicalAGP}, the exponent of the spectral function rapidly changes from $1/2$ to $1$ and further increase of disorder simply leads to increasing range of frequencies with $1/\omega$ asymptotic behavior, which after some possibly diffusive part is ultimately cutoff by system size at small frequencies. Our numerical results, at the present system sizes, do not support existence of a stable disorder dependent intermediate exponent. We note that this inverse frequency scaling of the spectral function was found earlier in random regular graph models~\cite{Kravtsov_RRG_2015, Kravtsov_Bethe_2018}.

As disorder increases beyond $W=2.5$ (bottom panel in Fig.~\ref{fig:spectralL16}, the slope of the spectral function decreases again. Simultaneously, a difference between typical and average spectral functions starts to develop signaling ergodicity is broken. The typical spectral function (right panel) stays closer to the $1/\omega$ scaling with a rapidly decreasing coefficient as a function of $W$. This change in behavior occurs when the Thouless energy $\omega_{\rm Th}$ becomes comparable the typical level spacing~\cite{NoteFits}. From conservation of the spectral weight~\eqref{eq:sum_rule} we conclude that the system goes into the localized (MBL) regime. This conclusion is further supported by analyzing the conserved part of the local magnetization (see Fig.~\ref{fig:conservedZ}). Remarkably, even at strongest disorder shown $W=15$ there is a clear tail down to zero frequency, which is well below the mean level spacing. So the localized phase consists of a mixture of localized and delocalized degrees of freedom. It is this low frequency tail, which is responsible for a noticeable system size dependence of the fidelity susceptibility even at strong disorder (see Fig.~\ref{fig:typicalAGP})

\section*{Dynamical obstruction}

Let us now estimate the Thouless energy, and argue about its fate, as well as about the fate of the localization transition as we increase the system size. First let us note that, as disorder increases, crossing a threshold of the order of $W\sim 1$, the onset of $1/\omega$ scaling of the spectral function starts at a high frequency $\omega_{\rm uv}\sim 0.1/W$, which is system size independent. This observation immediately implies that, in this scaling regime, $|f(\omega) |^2\approx C/\omega$ with the prefactor $C$ independent of the system size. Numerically we find that $C$ is inversely proportional do the disorder strength such that in this regime
\be
|f(\omega)|^2\approx { 0.0179 \over \, W \omega},
\ee
where the constant was extracted from the data in Fig.~\ref{fig:spectralL16}~\cite{NoteFits}. One can check that frequencies above this scale: $\omega>\omega_{\rm uv}$ give small and approximately disorder independent contribution to the spectral weight. Therefore it is the subdiffusive $1/\omega$ scaling regime of the spectral function, which dominates the total spectral weight. Because the integral of $1/\omega$ diverges, and in order to satisfy the sum rule~\eqref{eq:sum_rule}, there must be a low frequency scale $\omega_{\rm th}$, where the spectral function either saturates or potentially crosses over to a lower power of the inverse frequency corresponding to faster dynamics, for example diffusion. Assuming that $L$ is very large, this scale can be found from:
\[
2\int_{\omega_{\rm th}}^{\omega_{\rm uv}}  { 0.0179 d\omega\over W \omega}\approx {1\over 4}\quad \rightarrow \quad \omega_{\rm th}\approx \omega_{\rm uv} \mathrm e^{-\alpha W},
\]
with $\alpha\approx 6.98$. Note that the factor of 2 multiplying the integral stems from the equal contribution coming from negative frequencies. As we start decreasing $L$ at some point this scale hits the level spacing, after which the spectral weight has to condense and the system goes into the localized MBL regime. This happens when
\[
\omega_{\rm uv} \mathrm e^{-\alpha W}\approx \mathrm \Omega_{\rm uv} \mathrm e^{-L \log 2}, 
\]
where $\Omega_{\rm uv}$ is set by the many-body bandwidth and hence is bounded by $\Omega_{\rm uv} < cL$. This gives the approximate relation between the critical disorder strength and the system size at the transition point:
\be
\label{eq:W_ast}
W^\ast(L) \approx {L\log 2\over \alpha}-{\log (\Omega_{\rm uv}/\omega_{\rm uv})\over \alpha}
\ee
At large $L$ the critical disorder increases linearly in $L$, since $\log (\Omega_{\rm uv}/\omega_{\rm uv})$ only generates logarithmic corrections, such that $W^\ast\approx 0.099 L+c$. The latter is consistent with the prediction of Ref.~\cite{suntajs2019quantum}, in which the critical disorder strength was found to shift like $W^\ast \approx 0.098 L+c$ based on a scaling analysis of a many-body version of the logarithmic Thouless conductance.

Note that we extract a lower bound on the Thouless energy by using the f-sum rule, since we assumed all of the spectral weight to be in the $1/\omega$ regime of the spectral function. Numerically we find for $L=16$ that the UV part of the spectral weight, defined as $\int_{\omega_{\rm uv}}^\infty d\omega |f^2(\omega)|,\; \omega_{\rm uv}\approx 0.1/W$ takes up a significant fraction of the spectral weight, slowly reducing with $W$ from $74\%$ at $W=1.5$ to $46\%$ at $W=2.5$. It is expected that as $W$ increases this UV part goes down to zero and Eq.~\eqref{eq:W_ast} is asymptotically exact. But for available system sizes, in which the range of accessible disorder strengths showing $1/\omega$ scaling regime is limited, the constant $\alpha$ is significantly renormalized by this spectral weight reduction. If we instead identify the Thouless energy as the point of maximal curvature on~Fig.~\ref{fig:spectralL16}, we extract a different estimate for the exponent $\alpha$: $\alpha' \approx 3.43$ for a system of $L=16$ (see the inset in Fig.~\ref{fig:spectralL16}).  This is approximately half the maximal value extracted from the sum rule~\eqref{eq:W_ast} and it is in good agreement with the $\alpha''=1/0.29\approx 3.45$ estimated in~\cite{suntajs2019quantum}. It should be noted that the Thouless energy has previously been numerically extracted  from the spectral function (up to $L=20$) by Serbyn et al.~\cite{PhysRevB.96.104201}. Whereas the data is consistent with ours, it is analyzed differently and the authors had a different interpretation, supporting the Griffiths phase picture on the ergodic side, leading up to the critical fan where the Thouless energy decays exponentially with the system size $\omega_{\rm th}\propto \exp(-\kappa L)$. 

Let us end this section by noting that the commonly accepted Griffiths picture, with a critical point at finite disorder, is not internally consistent. The argument requires only three inputs: locality, conservation of energy and bounded norm of local operators. Consider a Griffiths scenario~\cite{kratiekreview} where local correlations, of a conserved operator like magnetization or simply an operator coupled to the energy, decay in time like $C(t)\sim t^{-1/z}$, where the dynamical exponent $z$ depends on disorder $W$. In the Griffiths picture, $z$ would be proportional to the correlation length $z\sim \xi(W)$. Since energy is conserved, decaying correlations imply spreading of this magnetization in space like $\Delta x\sim t^{1/z}$ and the spreading stops at the Thouless time when the dynamics detects the boundaries, i.e. the Thouless energy obeys $\omega_{\rm th}\sim L^{-z}$. In order for ergodicity to be broken one needs the Thouless energy to reach the level spacing, which implies $z  \gtrsim L/\log L$. In the Griffiths picture, this condition is equivalent to saying that the correlation length $\xi(W)>L$ has to exceed the system size, and as such the system is critical. In frequency space the spectral function for general subdiffusion with dynamical exponent $z$ simply reads 
\begin{equation*}
|f(\omega)|^2=\frac{C}{\omega^{1-1/z}},
\end{equation*}
where $C=|f(\omega=1)|^2$ is defined by the UV spectral weight. Note that, thus far, this scenario would also yield exponential enhancement of the susceptibility in the critical region. However, it suffers from the same problem as before, namely the total amount of spectral weight must be conserved. The latter directly implies 
\begin{equation*}
Cz \leq 1.
\end{equation*}
This inequality can be satisfied for any finite $z$, but for the system to become critical $z$ has to diverge with $L$ implying $C$ has to decay to zero at least like $\log(L)/L$. Not only is this inconsistent with the numerical data (see Fig.~\ref{fig:spectralLscaling}), it is simply impossible by locality. Recall that $C$ is the UV spectral weight at the onset of the asymptotic hydrodynamic scaling behavior. As such it can not possibly know about the system size $L$. Of course the spectral weight can be a function of disorder $W$, and while we argued that $C\sim 1/W$ such that $W^\ast \sim L$ faster decay will simply result in a slower drift of the critical point.

This argument is consistent with the observed existence of the localization transition in the non-local models like the random regular graphs~\cite{Kravtsov_RRG_2015, Kravtsov_Bethe_2018}. Note that in principle one can imagine other scenarios, where the exponent $z$ for example would be frequency dependent, approaching unity at the critical point only in the limit $\omega\to 0$. Such a scenario could correspond to logarithmic corrections to the $1/\omega$ scaling or a scenario with extra time scales. However, such scenarios are inconsistent with the Griffith physics as they imply that, already in the ergodic side, the system's dynamics is getting slower and slower with time instead of accelerating as the system ``realizes'' that it is in the ergodic regime. Also, such scenarios are inconsistent with existing numerical results both reported here in Fig.~\ref{fig:spectralL16} and in other works~\cite{PhysRevB.100.214313,PhysRevLett.118.196801}. The only remaining possibility to satisfy the spectral sum rule with the system size independent $C(W)$ at $z\to\infty$ is when $C(W)\to 0$ at some critical disorder $W_c$. This scenario corresponds to opening a spectral gap in the system. Such a spectral gap is present in normal Bethe-ansatz integrable models~\cite{pandey2020chaos}, suppressing the exponential divergence of the fidelity with the system size. However, this gap usually exists only for special classes of perturbations, which keep the system integrable. There is no numerical indication that a spectral gap opens up in disordered systems.

\section{Fidelity susceptibility revisited}

For the sake of completeness we now revisit the fidelity susceptibility, as there are a few more interesting points to make. From Fig.~\ref{fig:typicalAGP} we clearly discern three different scaling regimes of the typical fidelity susceptibility denoted by three different colors. These regimes can be loosely labeled as i) Completely ergodic or ETH type, ii) Glassy with slow relaxation and iii) localized. In this section we will analyze these three regimes in more detail.

\begin{figure}[ht]
	\centering
	\includegraphics[width= 0.48\textwidth]{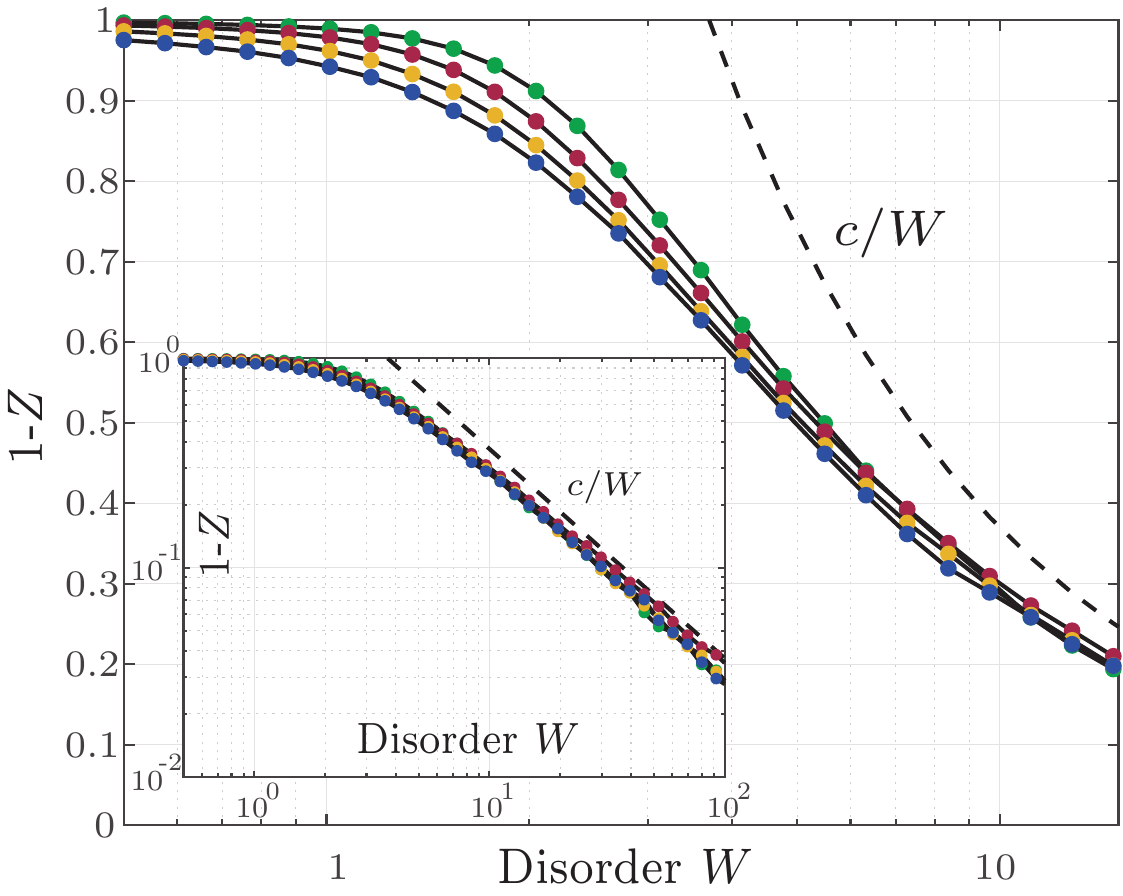}
	\caption{\textbf{Conserved Z-magnetization}. Conserved part of the local Z-magnetization defined in Eq.~\eqref{eq:Z_def} as a function of disorder for $L=10,12, 14, 16$ (yellow to blue). The main plot shows the magnetization in the log-linear scale and the inset does in the log-log scale. The dashed line shows the prediction of perturbation theory in $1/W$: $Z=1-c/W$. }
\label{fig:conservedZ}
\end{figure}

First we analyze the total conserved part of the local Z-magnetization, which is directly related to the integral of the spectral function (see Eq.~\eqref{eq:sum_rule}). We define the latter as 
\be
\label{eq:Z_def}
Z=4 \exp\left[ \langle \langle {\log(\langle n |S_l^z |n\rangle^2)\rangle\rangle}\right]
\ee
We choose the normalization such that $Z=1$ in the fully localized regime and $Z~\sim 2^{-L}$ in the ergodic regime in the zero magnetization sector. To avoid effects of the tail of the distribution of magnetization we first take its logarithm, then average it over different eigenstates and disorder realizations and then exponentiate. In Fig.~\ref{fig:conservedZ} we show $1-Z$, measuring the non-conserved part of the local magnetization. It is evident that as disorder increases the magnetization starts to localize. As we increase the system size the transition to the localized regime happens for larger and larger values of disorder and the observed drift is consistent with the drift of the peak in the fidelity susceptibility (see Fig.~\ref{fig:typicalAGP}). At very large $W$ all curves collapse into one and thus the conserved magnetization becomes insensitive to the system size. The black line is shown for reference representing the leading perturbative correction to the infinite disorder limit.

\begin{figure}[ht]
	\centering
	\includegraphics[width= 0.48\textwidth]{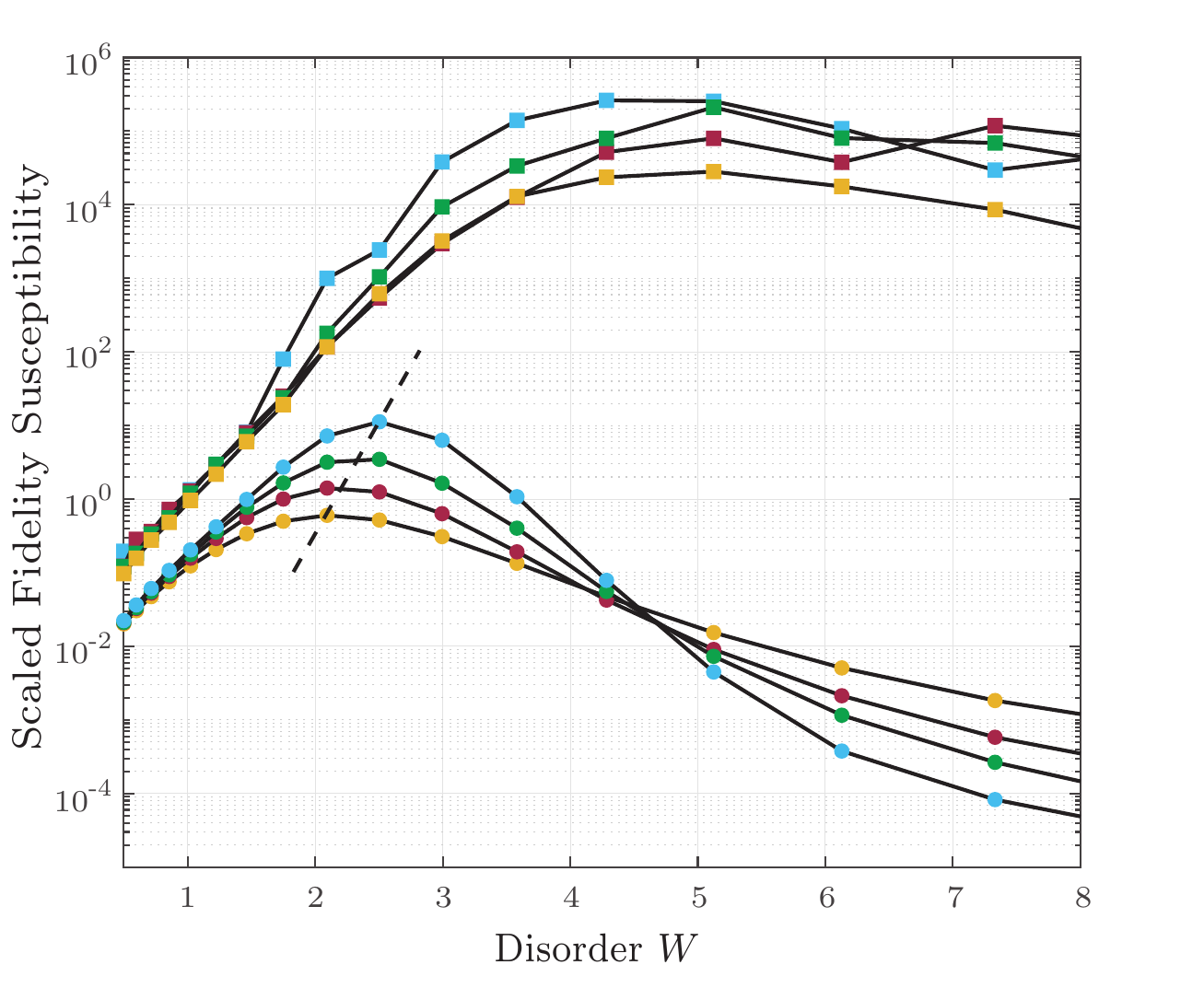}
	\caption{\textbf{Ergodicity breaking.} The scaled fidelity susceptibility is shown as a function of disorder for chains of length $L=12,14,16,18$ (yellow, red, green, blue). In contrast to figure~\ref{fig:typicalAGP}, we show both the typical (circles) and mean (squares). Moreover, by using a linear scale for the disorder $W$ it becomes directly apparent that both the typical and the mean grow exponentially at small disorder. After the peak in the typical susceptibility the mean susceptibility keeps growing, resulting in an exponentially large separation between the two. The dashed line shows the predicted drift of the maximum: $\chi_{\rm max}=c\exp[\alpha W^\ast]$, where $W^\ast=L \log(2)/\alpha$ (see Eq.~\eqref{eq:W_ast})}
\label{fig:colorplotAGP}
\end{figure}

Next in Fig.~\ref{fig:colorplotAGP} we show the mean scaled fidelity susceptibility $\bar \chi/2^L\equiv \langle\langle \chi_n \rangle\rangle/2^L$ and the typical one $\exp[\zeta]$ versus disorder. The plot of the latter (solid lines black with circles) reproduce the data from Fig.~\ref{fig:typicalAGP} except disorder is now shown in the linear scale. In the ergodic regime at low disorder the mean and the typical susceptibilities are parallel to each other with only a constant offset between them, which comes from small Gaussian fluctuations of $\chi_n$ around the mean. As disorder increases the typical susceptibility reaches the maximum and goes down, while $\bar\chi$ keeps growing until it saturates at large $W$. As we discussed earlier the saturated value is determined by rare resonances in Eq.~\eqref{eq:chi_n}, where $|E_n-E_{n\pm 1}|\ll 2^{-L}$. Such resonances inevitably occur due to absence of level repulsion (see also Fig.~\ref{fig:distributionAGPexponent}). As a result the mean fidelity susceptibility fluctuates much more at strong disorder than the typical one.

\begin{figure}[ht]
	\centering
	\includegraphics[width= 0.48\textwidth]{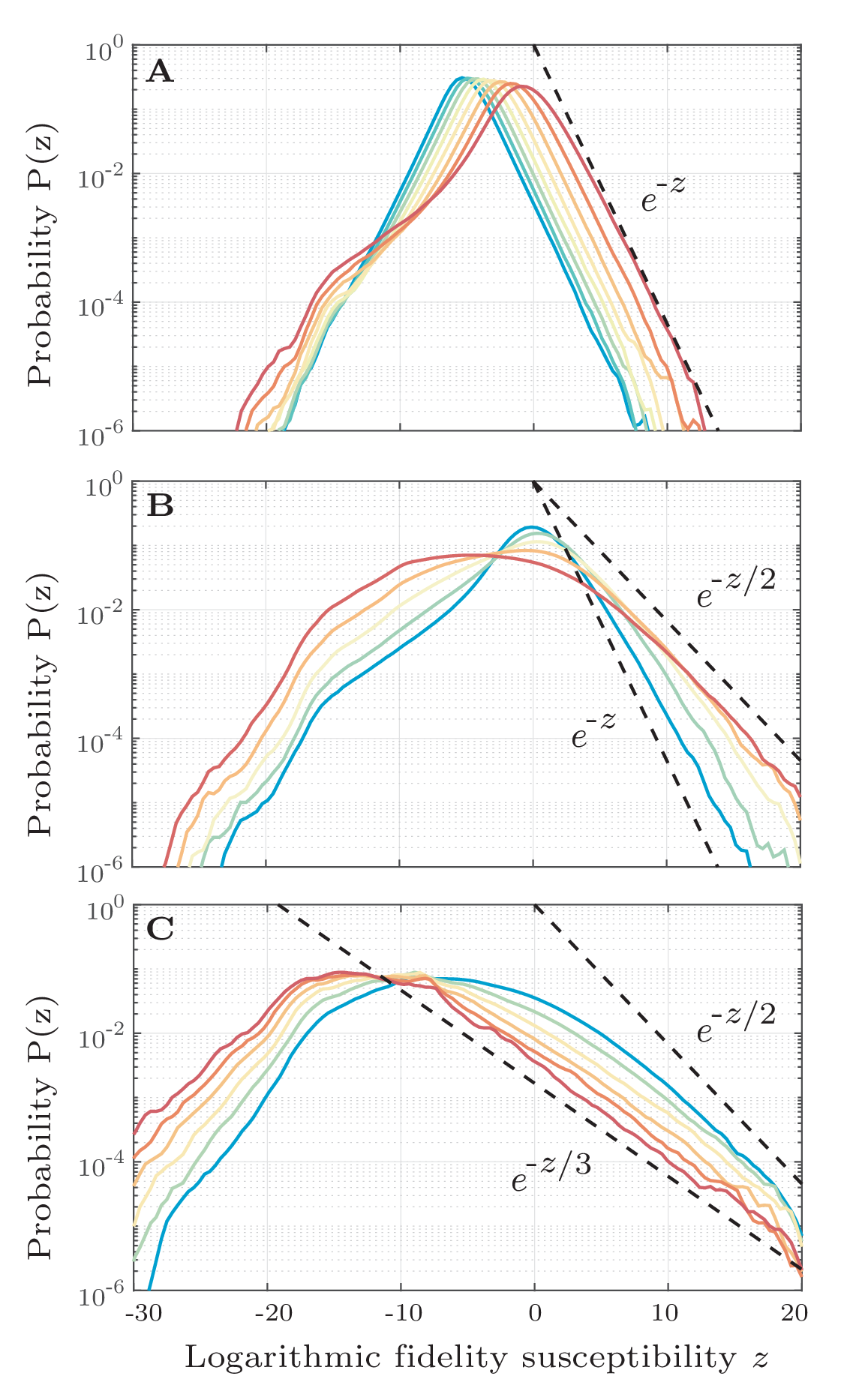}
	\caption{\textbf{AGP distribution}.  The distribution of the fidelity susceptibility is shown for a chain of $L=16$. The make the structure more apparent we have divided disorder range in 3 intervals. {\bf Panel A} Shows the disorder increasing from $W=0.5$ (blue) to $W=1.75$ (red) {\bf Panel B} Shows the disorder increasing from $W=2.1$ (blue) to $W=4.3$ (red) and {\bf Panel C} Shows the disorder increasing from $W=5$ (blue) to $W=15$ (red). The disorder is logaritmically equally spaced. Various exponentials are shown by dashed lines for comparison. The different panels are also labeled on Fig.~\ref{fig:distributionAGPexponent}.}
\label{fig:distributionAGP}
\end{figure}

\begin{figure}[ht]
	\centering
	\includegraphics[width= 0.48\textwidth]{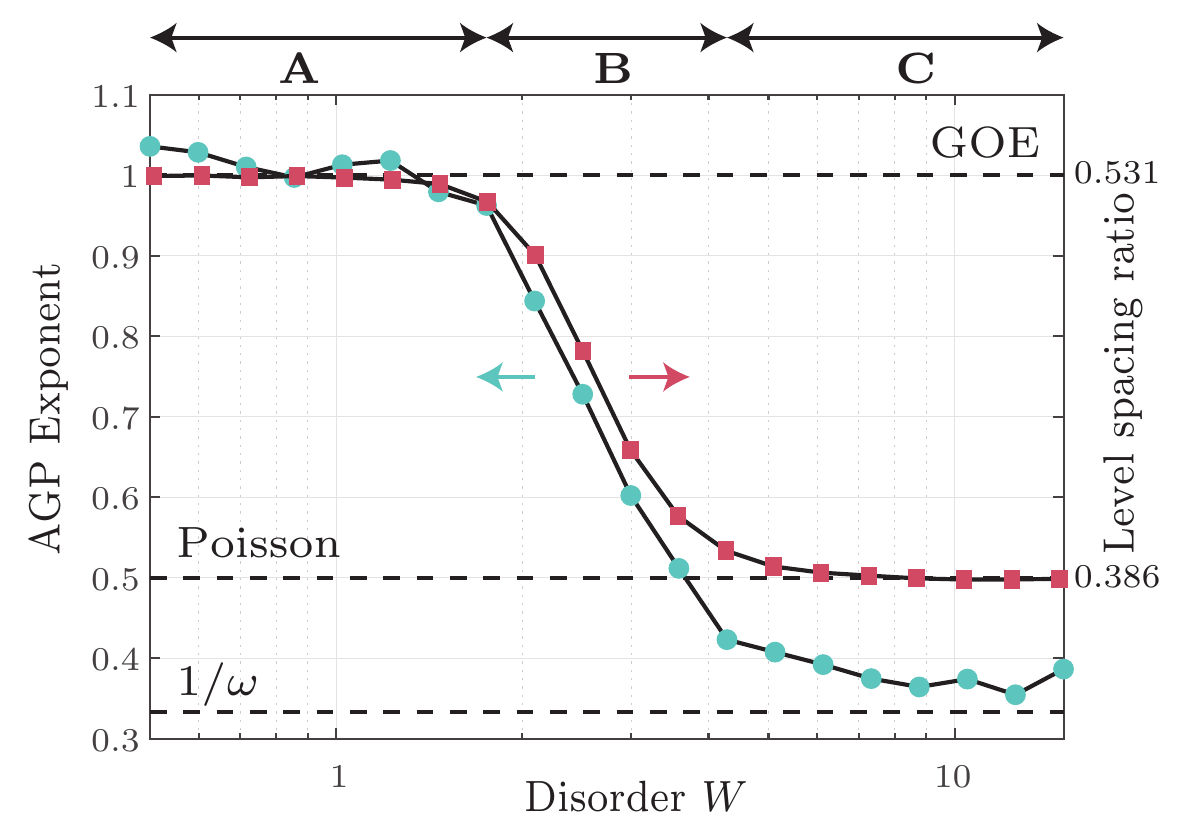}
	\caption{\textbf{AGP exponent}. The tail of the distribution of the logarithmic fidelity susceptibility decays with a characteristic exponent. This figure shows the exponent extracted from the distributions shown in Fig.~\ref{fig:distributionAGP} where the different panels are indicated on the top.  The fitted exponent is shown by the blue circles. For comparison, the red squares show the level spacing ratio. }
\label{fig:distributionAGPexponent}
\end{figure}

One can extract additional information about the system by analyzing the tail of the probability distribution of $\chi_n$ as it was first discussed in Ref.~\cite{pandey2020chaos}. Both in ETH and localized regimes this tail is determined by anomalously small nearest energy differences $|E_n-E_{n+1}|$ and thus contains additional information about the level statistics. For random matrix ensembles the distribution has been derived in~\cite{Sierant_2019} and a numerical survey on disordered systems has been presented in Ref.~\cite{Seriant19MBL}. To avoid exponential factors, it is more convenient to deal with the distribution of $\zeta_n\equiv \log(\chi_n)$. The probability that $\zeta_n=z$ at large values of $z$ can be estimated as
\[
{\rm P}(z)\equiv {\rm Pr}(\zeta_n=z)\approx {\rm Pr}\left(\log(|f^2(s)|/s^2)=z\right),
\]
where $s$ is the smallest of $|E_n-E_{n+1}|$ and $|E_n-E_{n-1}|$. Assuming that $|f(s)|^2=c |s|^{-\alpha}$ at small $s$ such that $s=~c\exp[-z/(2+\alpha)]$, we find
\[
{\rm P}(z)\propto \exp\left[-\frac{z}{2+\alpha} \right]{\rm Q}\left(c \exp\left[- {z\over 2+\alpha}\right]\right) ,
\]
where ${\rm Q}(s)$ is the probability to have the level spacing $s$. At small values of $s$, corresponding to large $z$, this probability usually takes a universal power law form: ${\rm Q}(s)\propto s^\beta$ with $\beta=0$ indicating Poisson statistics, $\beta>0$ corresponding to the level repulsion and $\beta<0$ corresponding to the level attraction. Combining all these factors together one finds 
\be
\label{eq:agp_tail}
{\rm P}(z)\propto \exp\left[-z {1+\beta\over 2+\alpha}\right].
\ee
In the ETH (random matrix) regime we have $\beta=1$ and $\alpha=0$ such that ${\rm P}(z)\propto \exp[-z]$, in agreement with the RMT result~\cite{Sierant_2019}. In the localized regime, with no level repulsion ($\beta=0$) and a GOE random operator $\partial_\lambda H$ with no selection rules $\alpha=0$, we would find ${\rm P}(\zeta_n=z)\propto \exp[-z/2]$. Note that this asymptote gives the slowest decay of the tail of the distribution for random operators in the absence of level attraction. Any exponent smaller than $1/2$ for such operators would necessarily imply $\beta<0$, i.e. the levels attract. However, for (normalized) operators with diverging spectral function $|f^2(s)|$ at $s\to 0$ with a positive exponent $\alpha$ the exponent in the tail of the distribution can be lower than $1/2$. Instead it is bounded from below by $1/3$, i.e. ${\rm  P}(z)\propto \exp[-z/3]$, which is realized at $\beta=0$ and the maximum possible value of $\alpha=1$. Any exponent less than $1/3$ would necessarily imply level attraction irrespective of the operator $\partial_\lambda H$.

In Fig.~\ref{fig:distributionAGP} we show the distribution functions of the log-fidelity susceptibility $\zeta_n$: ${ P}(z)$ for different disorder realizations. We divided the disorder realizations in the three intervals. The first interval (panel A) corresponds to disorder increasing from $W=0.5$ to $W=1.75$. In this interval the level spacing distribution is well described by the Wigner-Dyson statistics with $\beta=1$ (see Fig.~\ref{fig:distributionAGPexponent}) and $\alpha=0$ as the spectral function is flat at  frequencies of the order of level spacing (Fig.~\ref{fig:spectralL16}). So from Eq.~\eqref{eq:agp_tail} we anticipate that ${\rm P}(z)\propto \exp[-z]$, which perfectly agrees with the numerical results. As we keep increasing disorder the system starts to lose level repulsion (Fig.~\ref{fig:distributionAGPexponent}) and accordingly the slope in the tail distribution of $\zeta_n$ goes down. Panel B shows the results for disorder increasing from $W=2.1$ to $W=4.3$, where this slope clearly changes between $1$ and $1/2$. As disorder gets even larger (panel C) the slope of the tail keeps going down until it saturates at a value close to $1/3$, clearly below the bound $1/2$ for random operators, which we discussed in the previous paragraph. This exponent can be explained either by level attraction $\beta<0$ or by divergence of the spectral function $\alpha>0$. An independent check on the level statistics shows that the latter explanation is correct, i.e. $\beta=0$ and $\alpha>0$. In Fig.~\ref{fig:distributionAGPexponent} we show the extracted exponent as a function of disorder and contrast it with the level spacing statistics. Clearly the two curves qualitatively follow each other very well. The exponent is a bit noisier because there are higher sample to sample fluctuations in the distribution tail. As we emphasized at large disorder the exponent goes below a naive Poisson ratio, highlighting exponential enhancement of the matrix elements of magnetization at small energy differences compared to the matrix elements of random operators.

It is remarkable that analyzing the tail of the fidelity susceptibility we come to the same conclusion as analyzing the spectral function. Namely a simple cartoon representation of the localized phase as a set of weakly coupled ergodic boxes seems to be  invalid. This representation was, for example, underlying the renormalization group treatment of the MBL transition~\cite{vosk2015RG,potterRG}. Instead at small frequencies there is an exponential enhancement of the matrix elements of local operators as the local magnetization is not special in this respect. It is this enhancement that prevents the system from localization in the thermodynamic limit. At this point it is worth nothing that, while phenomenological RG's~\cite{vosk2015RG,potterRG} support the Griffiths phase, they are not consistent in that they do not predict the associated enhancement of the susceptibility. However, it is the same susceptibility which actually controls the RG. More microscopic constructions, such as those in~\cite{thiery2017microscopically}, are also controlled by the fidelity susceptibility and similar issues could emerge there as well.


\section{Conlusion}

We analyzed the sensitivity of eigenstates of a disordered one dimensional XXZ chain to small perturbations of a local longitudinal magnetic field. Specifically, we analyzed the typical fidelity susceptibility as a function of disorder strength and the system size. We found that this susceptibility exhibits a maximum near the localization transition, where the susceptibility is exponentially enhanced in the system size. We observe a significant drift of this peak to larger disorder with increasing system size, supporting the conclusions of recent work~\cite{Znidaric,suntajs2019quantum, suntajs2020transition,PhysRevLett.124.243601,kieferemmanouilidis2020absence}.

We further analyzed the low frequency dependence of the spectral function (auto-correlation function) of the local magnetization and found that for disorder larger than $W\sim 1$ it rapidly crosses over from a diffusive $|f(\omega)|^2\propto 1/\sqrt{\omega}$ to a subdiffusive regime with a universal exponent $|f(\omega)|^2\propto 1/\omega$. For the present system sizes, we did not find any evidence for existence of robust intermediate disorder-dependent exponents. We argued, based on simple physical principles, that this regime is inconsistent with localization in the thermodynamic limit because of conservation of the total spectral weight.

Finally we identified an issue with earlier analytic treatments of the MBL transition. Namely, the maximally chaotic critical region separating the ergodic and localized regimes is characterized by a strong divergence of the matrix elements of local operators at small energy differences of the order of the level spacing. This enhancement is not captured by phenomenlogical RG's and, according to our arguments, leads to instability of the localized phase at large system sizes. Despite finding evidence against localization in the thermodynamic limit, we confirmed existence of a very slow, nearly localized subdiffusive regime with logarithmic in time spreading of correlations.

The predictions of our work can be readily tested experimentally. In particular, the spectral function can be extracted as a Fourier transform of the connected part of the auto-correlation function of the local magnetization, which was measured already for over three decades in state of the art cold atom experiments~\cite{bloch2019mbl}. One can also extract it from transport measurements analyzing spreading of an initially localized spin or charge like it was done in Anderson localization experiments~\cite{billy2008Anderson}. In time, one might also consider extracting it using a quantum computer~\cite{sels2019quantum,vasilyev2020monitoring}. While extracting the very low frequency part of the spectral function might be challenging, the universal $1/\omega$ scaling starts at accessible frequencies and should be measurable.

\section*{Acknowledgements}
The authors thank Mohit Pandey for his collaboration during the early stages of this work. They further acknowledge inspiring discussions with Ehud Altman, Anushya Chandran, Soonwon Choi, Phil Crowley, Eugene Demler, David Huse, Frank Pollmann, Toma{\v{z}} Prosen, Marcos Rigol, Giuseppe De Tomasi, and Lev Vidmar. The authors also acknowledge useful feedback on the manuscript by Dima Abanin, Anatoly Dymarsky, Sarang Gopalakrishnan, Maksym Serbyn, and Michiel Wouters. A.P. was supported by NSF DMR-1813499 and AFOSR FA9550-16-1-0334. 

\begin{appendix}

\section{Appendix: Strong Disorder scaling}
\label{app:Scaling}

Here we will have a look at the behavior of the typical fidelity susceptibility $\zeta$ for $W \rightarrow \infty$. Recall, we are investigating the susceptibility with respect to changing the local magnetic field, such that 
\begin{equation}
\chi_n=  \sum_{m\neq n}{|\langle n | S^z_l |m\rangle|^2\over (E_n-E_m)^2},
\end{equation}
and $\zeta=\langle\langle\,\log(\chi_n)\,\rangle\rangle$. When $W \rightarrow \infty$ the Hamiltonian becomes diagonal in the $S^z$ basis, consequently the susceptibility tends to zero. As long as there are no (many-body) resonances, which do not affect the typical susceptibility anyway, we can do perturbation theory in $1/W$. To first order we find 
\begin{equation}
\chi^{(1)}_n= \frac{1}{W^2} \sum_{m\neq n} (s_l^n-s_l^m)^2 {|\langle z_n | H_{XX} |z_m\rangle|^2\over \left(E^{(0)}_n-E^{(0)}_m\right)^4},
\label{eq:firstorder}
\end{equation}
where $\left| z_n\right>$ are $z$-polarized product states, $s_l^n=\pm 1/2$ is the value of $S^z_l$ in this eigenstates, i.e. $S^z_l \left| z_n \right> =s_l^n \left| z_m \right>$, and the energy  $E_n^{(0)}=\sum_i h_i s_i^n$. Finally, the flip-flop Hamiltonian $H_{XX}$ is defined as
 \be
H_{XX}= \sum_j ( S^x_j S^x_{j+1}+S^y_j S^y_{j+1}).
\ee
It's directly clear from eq.~\eqref{eq:firstorder} that all contributions to the susceptibility in which $m$ and $n$ have the same value of the spin on site $l$ vanish. Since $H_{XX}$ simply swaps neighbouring spins, the only states that contribute to the sum in expression~\eqref{eq:firstorder} are those which have at least one of the neighbours of $l$ anti-aligned with $l$, i.e. 
\begin{equation}
\chi^{(1)}_n= \frac{1}{4W^2} \left[ \frac{(s_l^n-s_{l+1}^n)^2}{(h_l-h_{l+1})^4}+ \frac{(s_l^n-s_{l-1}^n)^2}{(h_l-h_{l-1})^4} \right].
\label{eq:firstorder2}
\end{equation}
Considering all 8 possible combinations of $s_{l}$ and $s_{l\pm1}$, there are four states for which one has a contribution of $1/4W^2\Delta h^4$, two states for which both neighbours contribute and we have $1/2W^2 \Delta h^4$, and two states states for which we get $0$. The latter, which are completely polarized states $\left| \cdots \uparrow \uparrow \uparrow \cdots \right>$ and $\left| \cdots \downarrow \downarrow \downarrow \cdots \right>$, need to be handled with care in computing the typical susceptibility $\zeta_n=\log \chi_n$, as this would diverge. For those states we need to consider higher order corrections in perturbation theory, since the first order contribution vanishes. Fortunately that is straightforward to do, as long as there are no resonances such that we can neglect renormalization of the denominators. 
To first order we had 
\begin{equation}
\zeta = \frac{3}{4} \log W^{-2} + \frac{1}{4} \log O(W^{-4})+B,
\end{equation}
where $B$ is some universal constant that is independent of $L,W$ and comes from the disorder average over $h$. In order for the polarized states to contribute in the next-order, they need move an anti-aligned spin to the center by an extra application of the flip-flip Hamiltonian $H_{XX}$. Consequently, either the left or right must have opposite spin to the polarized region, otherwise also the $O(1/W^4)$ contribution vanishes. Once again, we have a probability of $1/4$ for both of the ends to be aligned with the central piece. This argument can be applied ad infinitum, resulting in
\begin{eqnarray}
\zeta&=&  \frac{3}{4} \log W^{-2} + \frac{1}{4} \left( \frac{3}{4} \log W^{-4}+\frac{1}{4} \log O(W^{-6}) \right)+B \nonumber \\
&=& - \sum_{i=1}^\infty \frac{3}{4} \left( \frac{1}{4} \right)^{i-1} 2i \log W +B= -\frac{8}{3} \log W+B
\label{eq:logW}
\end{eqnarray}
The typical eigenstates susceptibility thus decays to zero like $W^{-8/3}$ for sufficiently strong disorder, in contrast to the average over eigenstates which behaves as $W^{-2}$. Finally, let us briefly look at finite size $L$ effects. Given that we are growing the polarized region from both sides in every order of perturbation theory, finite $L$ effects can be estimated by replacing the upper bound in~\eqref{eq:logW} by $L/2$. The latter would result in corrections of order $L2^{-L}$, hence convergence to the thermodynamic limit is fast. However, since the Hamiltonian under consideration conserves total magnetization, we have been primarily concerned with the zero-magnetization sector of the problem. 
The latter modifies the probabilities in a finite size system $L$ from our previous argument, as it becomes more and more unlikely to have have an all polarized region since the total magnetization ultimately needs to sum up to zero. 

\begin{figure}[h]
	\centering
	\includegraphics[width= 0.48\textwidth]{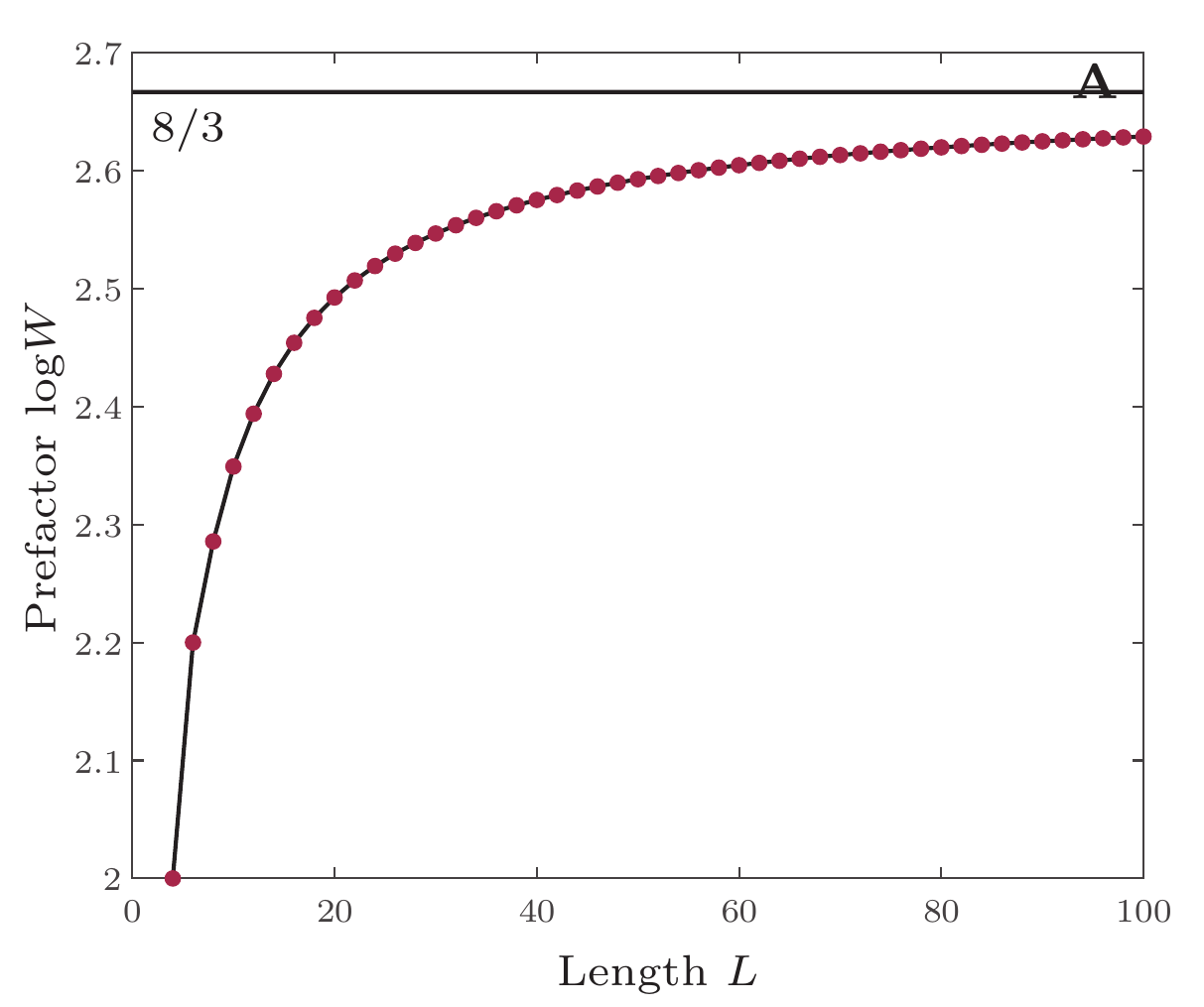}
	\includegraphics[width= 0.48\textwidth]{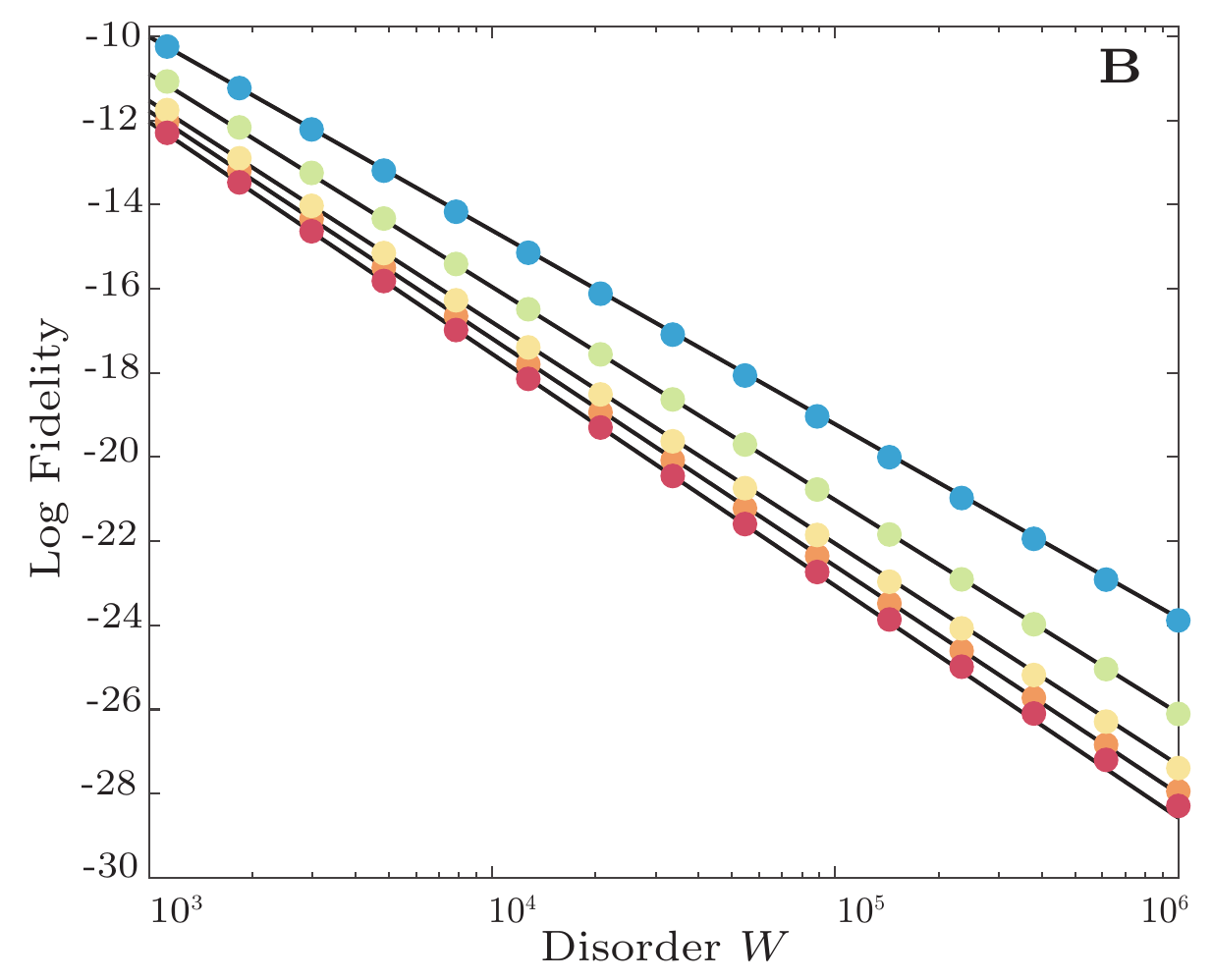}
	\caption{\textbf{$\log W$ prefactor}. Figure shows the behavior of the constant $c$ in the asymptotic behavior of the logarithmic fidelity susceptibility $\zeta=c \log W$. Panel A shows the numerical evaluation of expression~\eqref{eq:generalzetasum} and Panel B shows $\zeta$ extracted from exact diagonalization for systems sizes $L=4$ (blue) to $L=12$ (red) for very large disorder $W$. The black lines are fits of the form $\zeta=C \log W+B$, where $C$ is fixed from expression~\eqref{eq:generalzetasum} and only the offset $B$ is fitted. }
\label{fig:logprefactor}
\end{figure}

Consider the $i$th step, in which we are looking at the faith of a polarized region consisting of $2i-1$ spins. The probability that both the left and right neighbours have the same polarization is 
\begin{equation}
p_i= \frac{(L/2-(2i-1))}{L-(2i-1)}\frac{(L/2-2i)}{L-2i}.
\end{equation}
One can go through exactly the same argument as before, resulting in a generalization of expression~\eqref{eq:logW}:
\begin{equation}
\zeta= -\sum_{i=1}^{L/4} \prod_{j=0}^{i-1} \left[ p_j\right]\, (1-p_i)\, 2i \, \log W +B,
\label{eq:generalzetasum}
\end{equation}
where one should set $p_0=1$. The result is plotted in Fig.~\ref{fig:logprefactor}, showing how the exponent increases from $2$ to $8/3$ with increasing system size $L$. Due to the constraint, the convergence to the thermodynamic limit is only algebraic and corrections are of $O(1/L)$. Note that this causes a pronounced and rather counter-intuitive phenomenon for small systems, that in this asymptotic scaling regime the susceptibility of larger systems decays to zero faster than that of smaller systems. 

Finally note that, expression~\eqref{eq:firstorder} applies more generally in the MBL regime, as long as one simply replaces $s^n_l=\pm 1/2$ with the non-zero expectation values of the local $z$-magnetization in the dressed l-bit eigenbasis. It is easy to see that irrespective of the assumptions about the matrix elements this perturbation theory is exponentially divergent with the block size at small energy differences unless $(s^n_l-s^m)$ vanishes with the energy difference  $(E_n-E_m)$. As long as the latter is not true, this perturbation theory is unstable, leading to the spectral weight generation at small frequencies and decay of conserved magnetization. Further details of this instability will be reported elsewhere. 

\end{appendix}

\bibliography{ref_general} 

\begin{thebibliography}{63}
\bibitem [{Not()}]{NoteFits}%
  \BibitemOpen
  \href@noop {} {}\bibinfo {note} {The Thouless energy is extracted numerically
  by identifying the point of maximal curvature of the data on log-log scale.
  This procedure is robust up to $W \approx 1.5$. The coefficient $C$, for the
  asymptotic $C/W \omega$ behavior of the spectral function, is extracted by
  multplying the spectral function by $W\omega$ and extracting the value of the
  inflection point. This estimate smoothly goes from 0.0245 to 0.0179 when
  disorder changes from $W=1$ to $W=2.5$.}\BibitemShut {Stop}%
\end{thebibliography}

\end{document}